\documentclass[aps,prl,twocolumn,superscriptaddress,showpacs]{revtex4}
\usepackage{amssymb}
\usepackage{graphics}
\usepackage{amsmath}
\usepackage{subfigure}
\usepackage{times}

\def\tlfsx{TlFe$_{2-x}$Se$_2$}
\def\tlfs{TlFe$_{1.6}$Se$_2$}
\def\ktlfsx{(K,Tl)$_y$Fe$_{2-x}$Se$_2$}
\def\ktlfs{(K,Tl)$_y$Fe$_{1.6}$Se$_2$}
\def\kfs122{K$_y$Fe$_2$Se$_2$}
\def\kfsx22{K$_{1-x}$Fe$_2$Se$_2$}

\def\bfa122{BaFe$_2$As$_2$}
\def\fs11{FeSe}

\begin{document}

\preprint{}
\title{Block Spin Ground State and 3-Dimensionality of \ktlfs}

\author{Chao Cao}
  \affiliation{Condensed Matter Physics Group,
  Department of Physics, Hangzhou Normal University, Hangzhou 310036, China}

\author{Jianhui Dai}
  \affiliation{Condensed Matter Physics Group,
  Department of Physics, Hangzhou Normal University, Hangzhou 310036, China}
  \affiliation{Department of Physics, Zhejiang University, Hangzhou 310027, China}

\date{February 9, 2011}

\begin{abstract}
The magnetic properties and electronic structure of \ktlfs\ is
studied using first-principles calculations. The ground state is
checkerboard antiferromagnetically coupled blocks of the minimal
Fe$_4$ squares, with a large block spin moment $\sim 11.2 \mu_B$.
The magnetic interactions could be modelled with a simple spin model
involving both the inter- and intra-block, as well as the n.n. and
n.n.n. couplings. The calculations also suggest a metallic ground
state except for $y=0.8$ where a band gap $\sim 400-550$ meV opens,
showing an antiferromagnetic insulator ground state for
(K,Tl)$_{0.8}$Fe$_{1.6}$Se$_2$. The electronic structure of
the metallic \ktlfs~ is highly 3-dimensional with unique Fermi
surface structure and topology. These features indicate that the
Fe-vacancy ordering is crucial to the physical properties of
\ktlfsx.
\end{abstract}


\maketitle 

Superconductivity (SC) with moderate high transition temperatures
\cite{lofa_discovery,xhchen_nature_453_761,nlwang_prl_100_247002,zxzhao_epl_83_17002,hhwen_epl_82_17009,cwang_epl_83_67006}
has been observed in a broad family of the iron-based materials.
They are typically represented by the 1111-type LaFeAsO
\cite{lofa_discovery}, the 122-type BaFe$_2$As$_2$
\cite{Rotter-122}, the 111-type LiFeAs\cite{CQJin-111}, and the
11-type FeSe \cite{MKWu-11}. The parent compounds of these materials
show a universal strip-like (collinear) antiferromagnetic (SDW)
order \cite{pcdai_nature_453_899,PhysRevLett.101.257002} except for
the 11-type iron chalcogenides where the magnetic order is
bi-collinear\cite{PhysRevLett.102.247001,LuXiang-bi}. The magnetic
properties are closely related to a common two-dimensional Fe-atom
square lattice and the electronic structures are featured by the
cylinder-like hole and electron pockets around the $\Gamma$ and $M$
points respectively, with relatively weak dispersions alone the
$c$-axis. By electron or hole doping the Fermi surfaces evolve
smoothly in accordance with rigid band shift and the SC instability
is enhanced once the magnetic order is suppressed \cite{Paglione}.

Recently, a new family of the 122-type FeSe compounds
K$_y$Fe$_2$Se$_2$ \cite{ xlchen_prb_82_180520} and
(Tl,K)$_y$Fe$_{2-x}$Se$_2$ \cite{mhfang_1012,arxiv:1101.0462} have
been found to exhibit SC with transition temperatures $T_c\sim 30
K$. Moreover, the iron deficient compound (Tl,K)$_y$Fe$_{2-x}$Se$_2$
shows two remarkable features: (i) The SC (appears for $x\sim
0.12-0.3$, $y\sim 1$) is in proximity to an insulating phase ( for
relatively larger ~$x$, or $y\sim$
0.8)\cite{mhfang_1012,arxiv:1101.0462,arxiv:1101.0572}, and (ii) the
Fe-vacancies may exhibit some ordered
superstructures\cite{mhfang_1012}.  Early M\"{o}ssbauer experiment
for TlFe$_{2-x}$Se$_2$ \cite{Seidel} and recent transmission
electron microscopy on KFe$_{2-x}$Se$_2$ (for $x=0.4\sim 0.5$)
\cite{arxiv:1101.2059} provide clear evidence for the tetragonal and
orthorhombic superstructures in the FeSe layer. Previous
first-principles calculations suggested that the Fe-vacancy
orthorhombic superstructure could be stablized with an stripe-like
(collinear) AFM ground state in
(Tl,K)Fe$_{1.5}$Se$_2$\cite{arxiv:1012.6015,arxiv:1101.0533}.
The insulating behavior with an activation gap $\sim 60$ meV in
transport observed for $x\sim 0.5$ or around\cite{mhfang_1012} could
be attributed to a moderate large short-ranged Fe-3d electron
correlation, manifesting a possible Mott insulator driven by kinetic
energy reduction due to the ordered
Fe-vacancies\cite{arxiv:1101.0533}. The Mott-transition can be
indeed realized by a relatively smaller $U_c$ in a two-orbital model
with vacancy orderings \cite{YuZhuSi,ZhouZhang}.



\begin{figure*}[htp]
  \centering
  \subfigure[Geometry and Primitive cell]{
    \scalebox{0.2}{\includegraphics{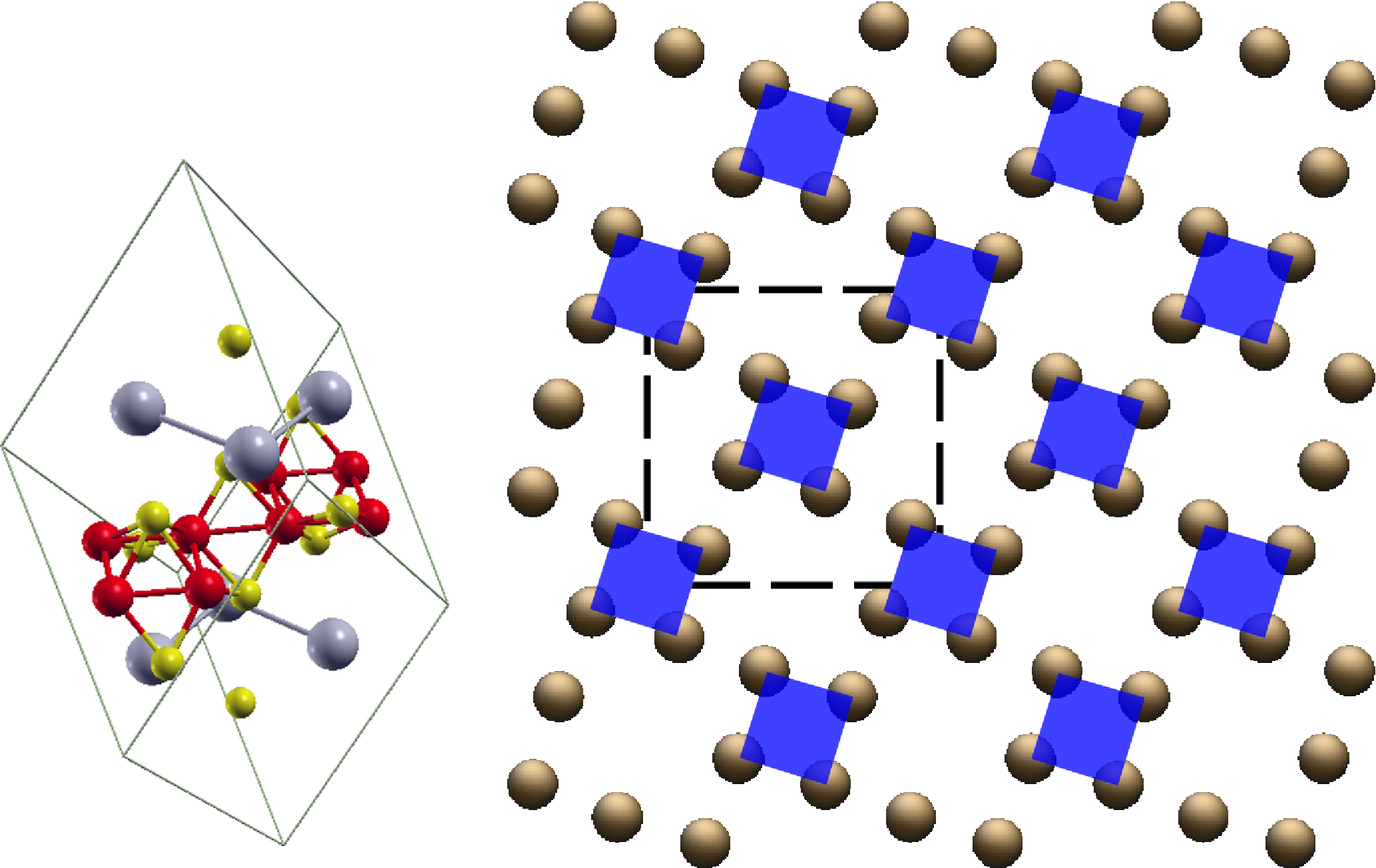}}
    \label{fig_structure_geometry}}
  \subfigure[Magnetic Couplings]{
    \scalebox{0.2}{\includegraphics{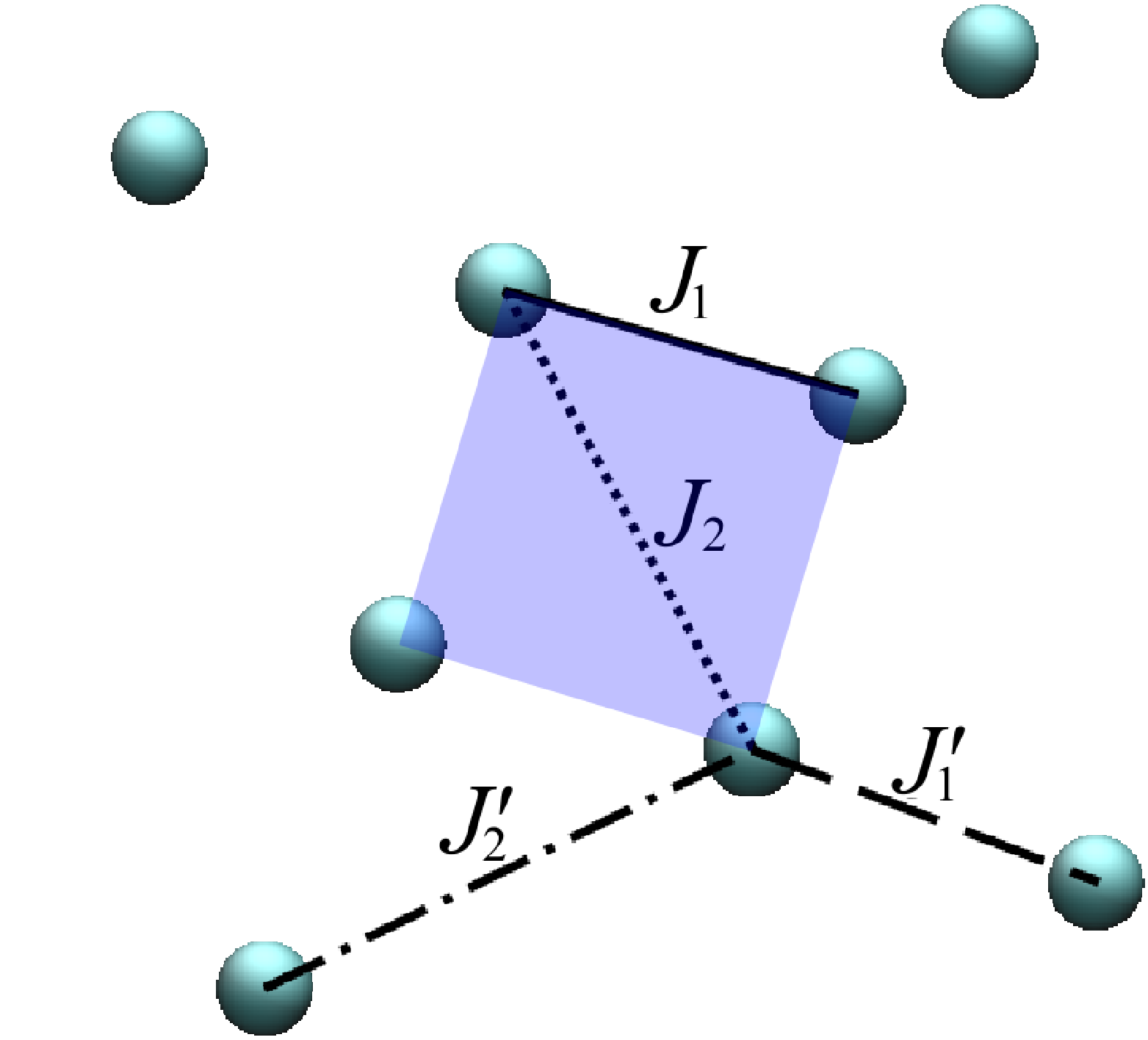}}
    \label{fig_mag_couple}}
  \newline
  \subfigure[AFM0]{
    \scalebox{0.12}{\includegraphics{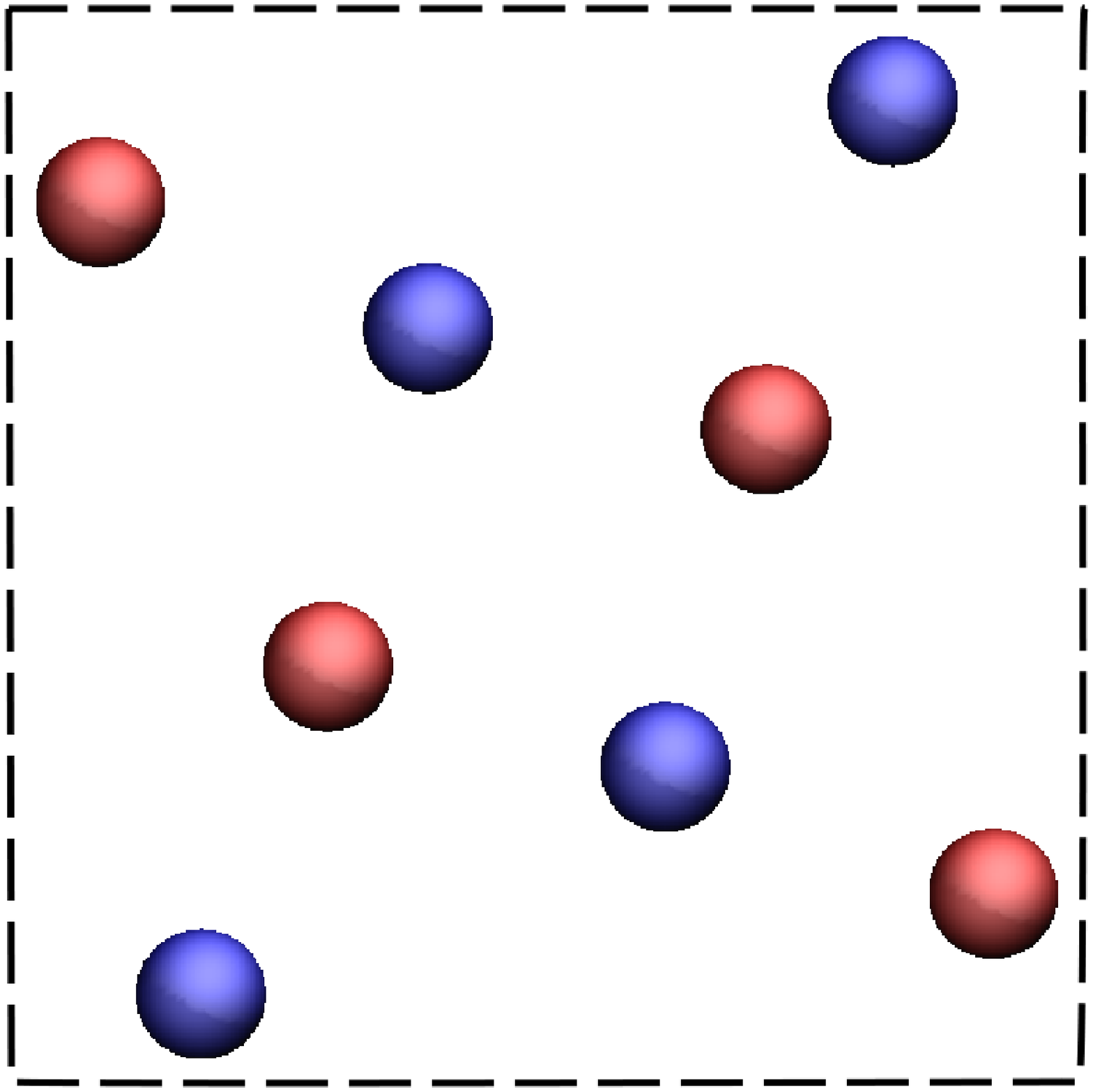}}
    \label{fig_afm0}}
  \subfigure[AFM1]{
    \scalebox{0.12}{\includegraphics{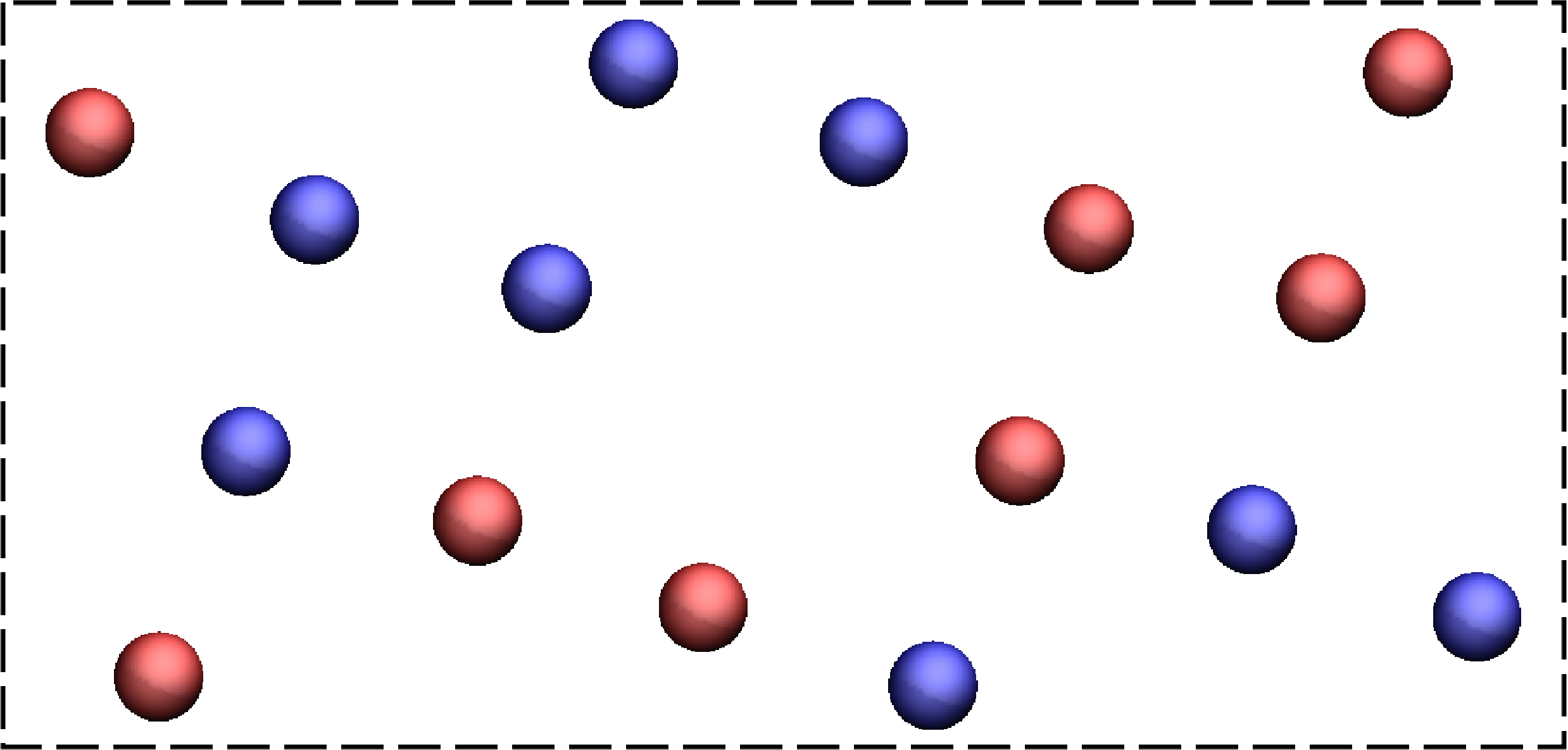}}
    \label{fig_afm1}}
  \subfigure[AFM2]{
    \scalebox{0.12}{\includegraphics{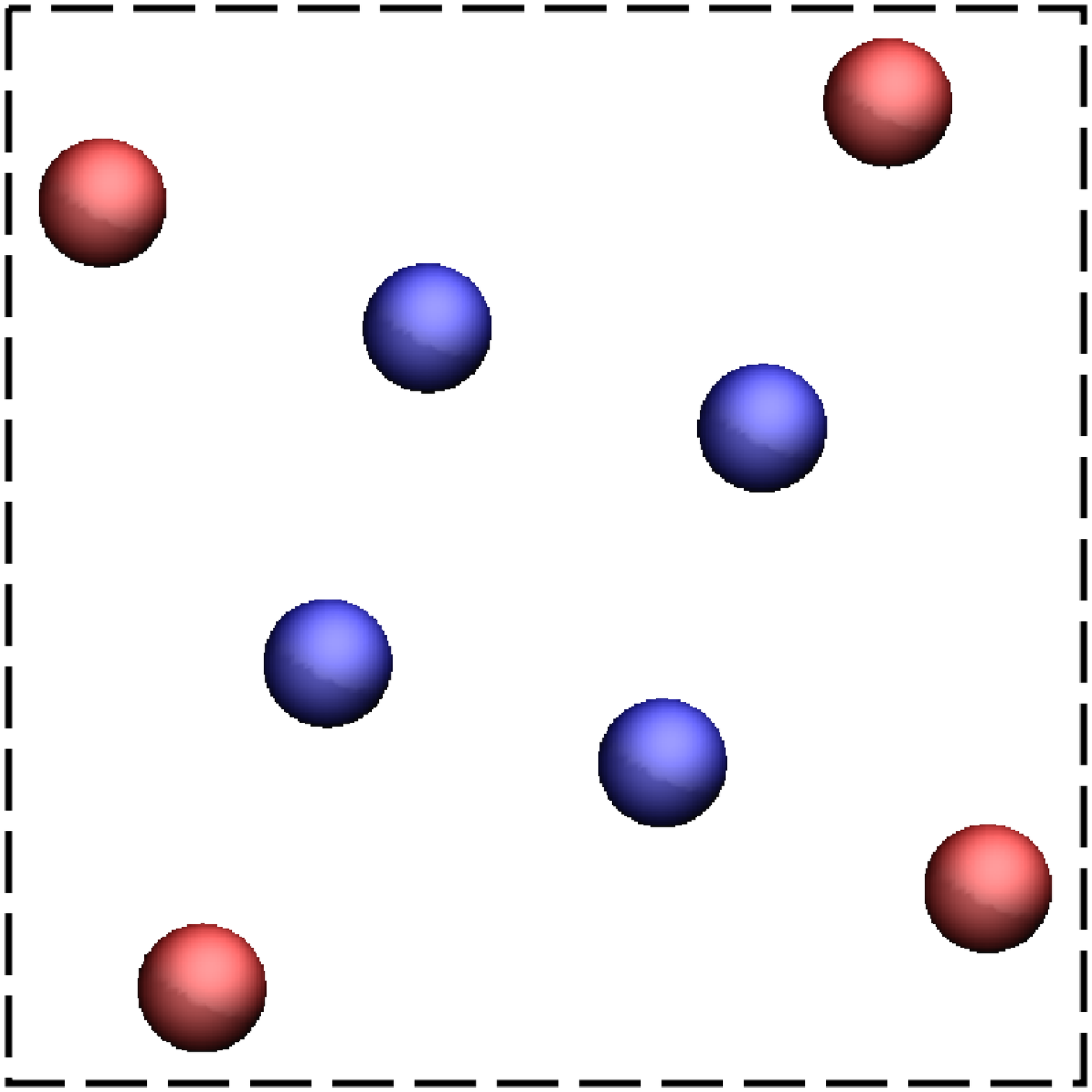}}
    \label{fig_afm2}}
  \newline
  \subfigure[AFM3]{
    \scalebox{0.12}{\includegraphics{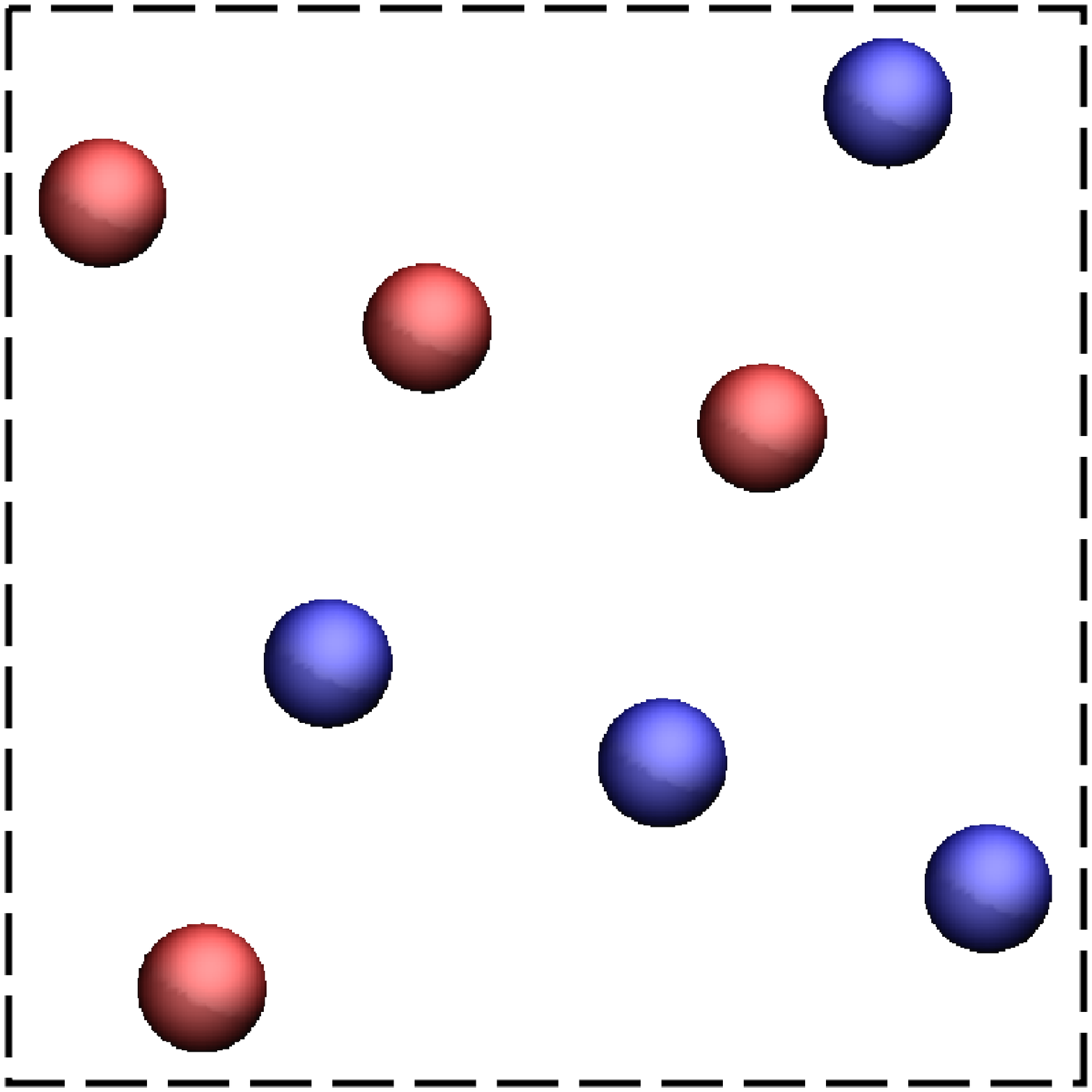}}
    \label{fig_afm3}}
  \subfigure[AFM4]{
    \scalebox{0.12}{\includegraphics{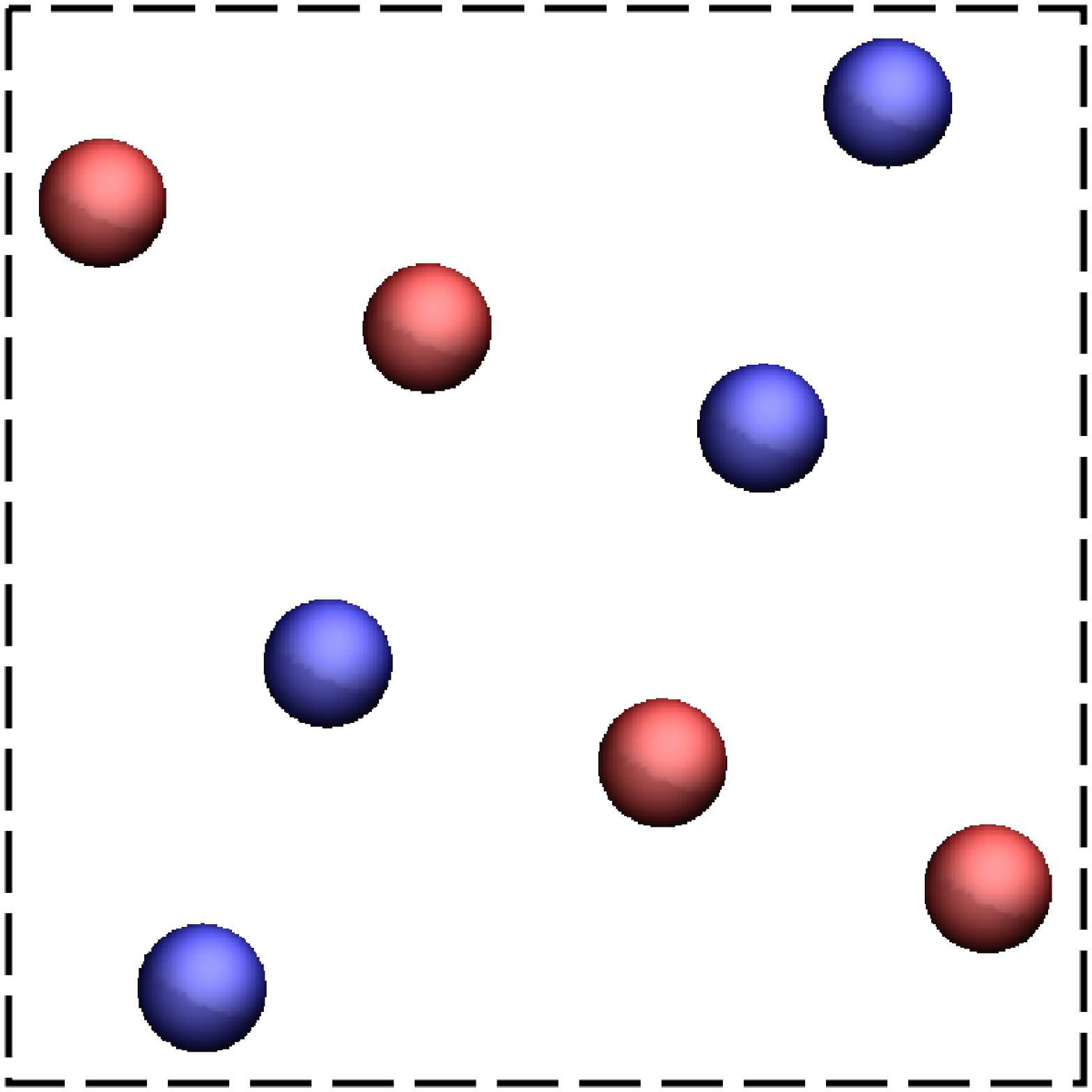}}
    \label{fig_afm4}}
  \subfigure[AFM5]{
    \scalebox{0.12}{\includegraphics{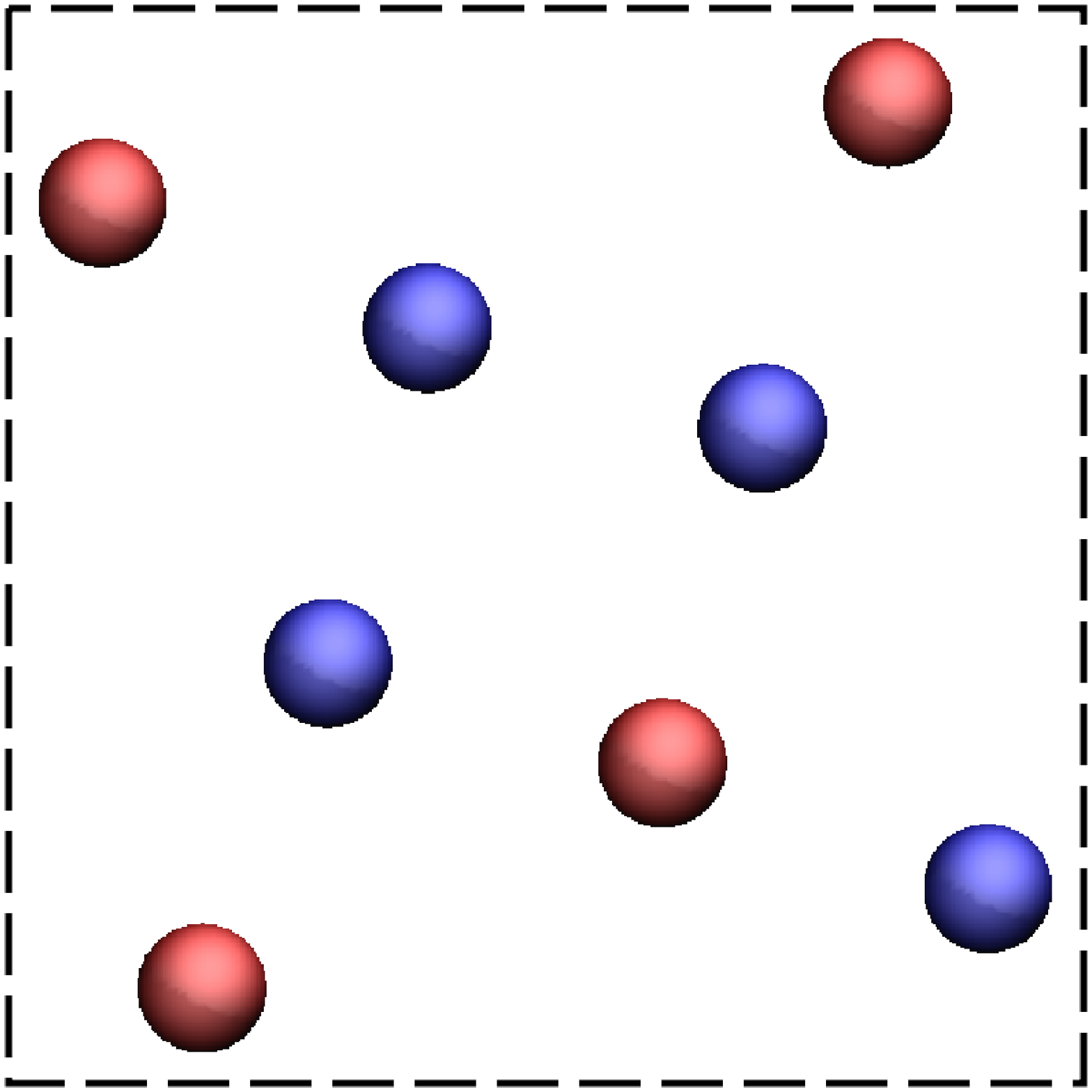}}
    \label{fig_afm5}}
  \subfigure[AFM6]{
    \scalebox{0.12}{\includegraphics{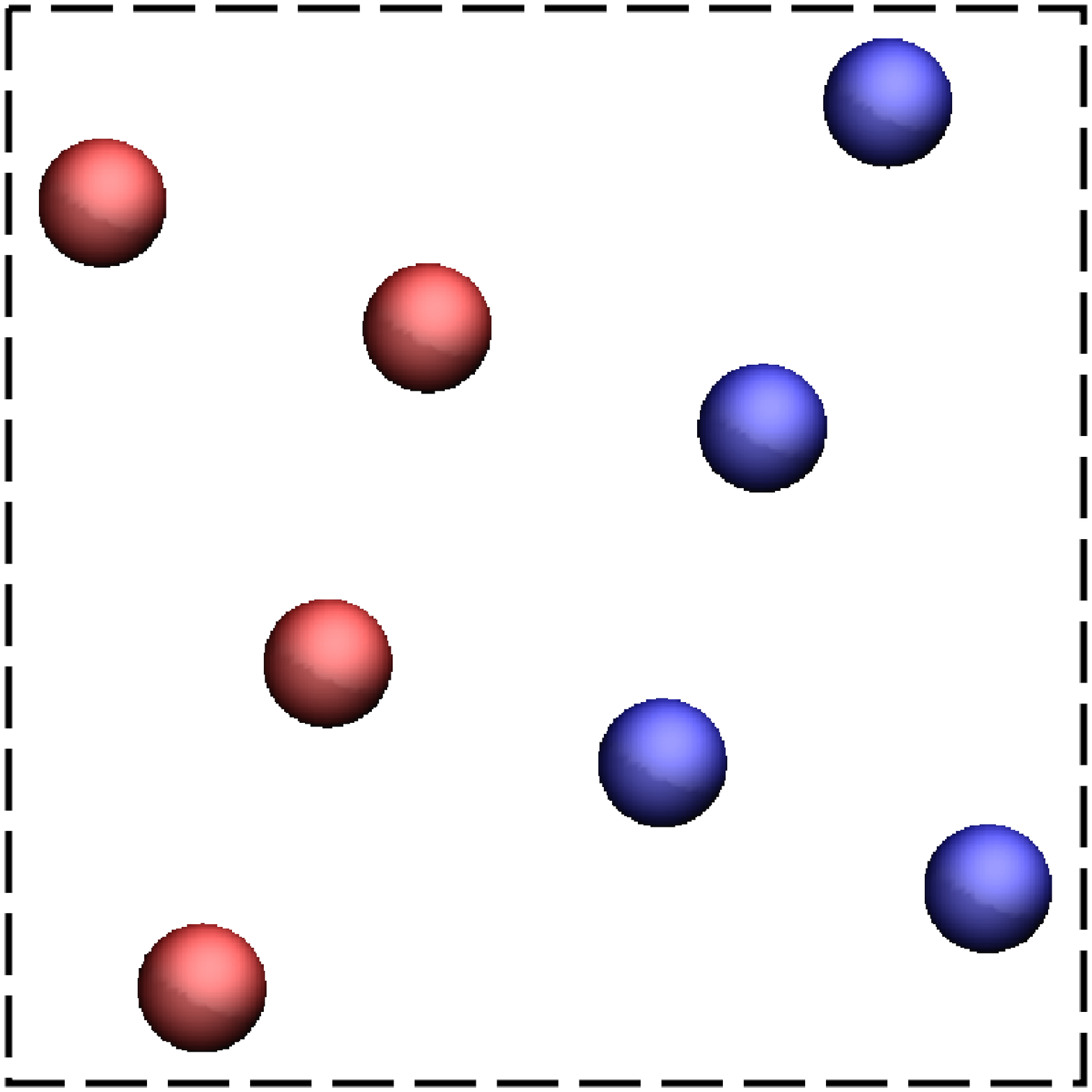}}
    \label{fig_afm6}}
  \subfigure[AFM7]{
    \scalebox{0.12}{\includegraphics{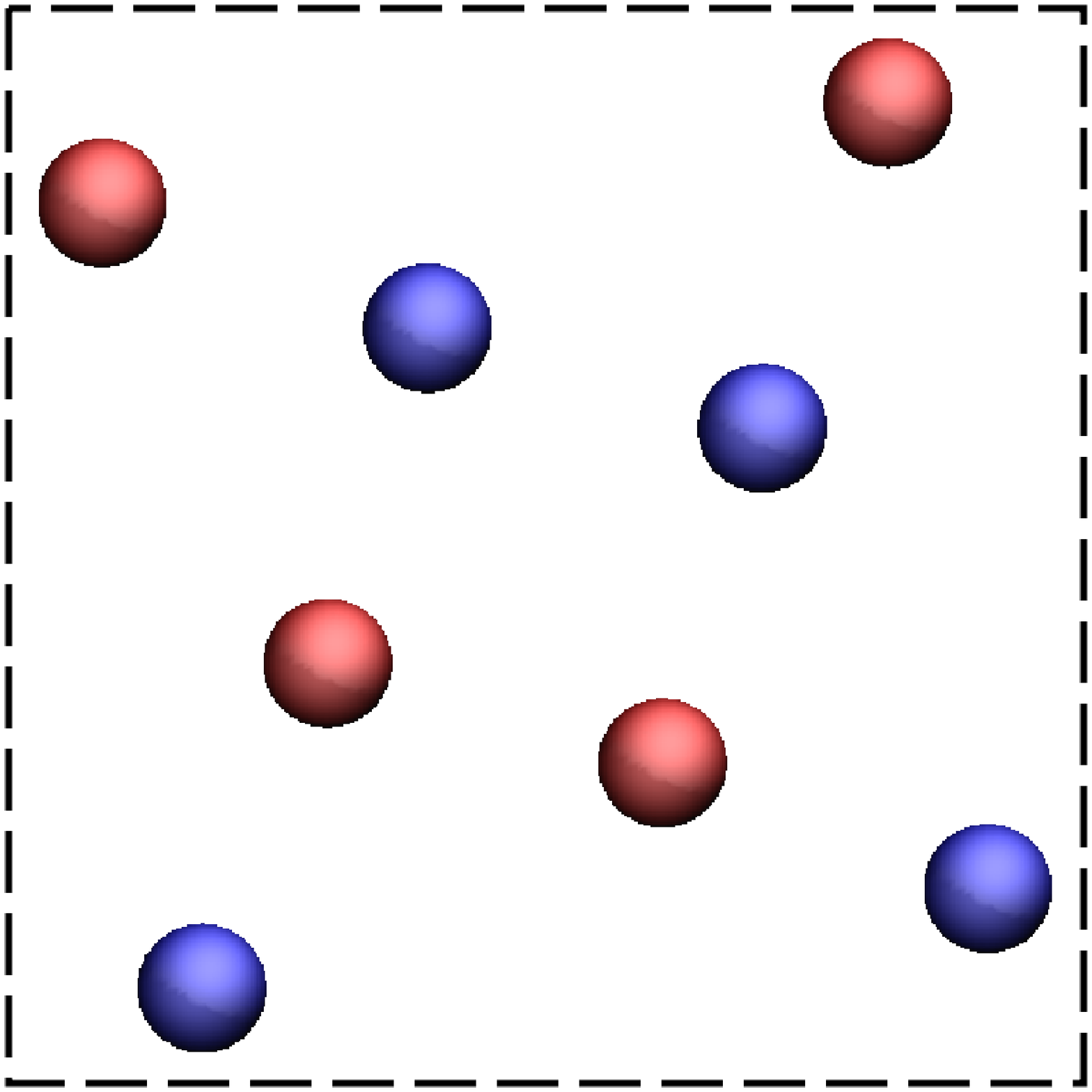}}
    \label{fig_afm7}}
  \caption{\ref{fig_structure_geometry}
The geometry and primitive cell from top-view. The shaded (blue)
region indicates the fundamental block with the four Fe-atoms at the
corners. The area encircled by the dashed line
for the AFM2 configuration (the ground state) is 74.75 \AA$^2$.
The periodic boundary condition for such blocks extended
over the whole lattice is imposed. \ref{fig_mag_couple} The proposed
magnetic couplings. ($J_1$, $J_2$) and ($J'_1$, $J'_2$) represent the
intra-block and inter-block (n.n., n.n.n.) couplings, respectively.
\ref{fig_afm0} to \ref{fig_afm7} Various magnetic configurations.
The red/blue atoms indicate the Fe-atoms with positive/negative
total magnetic moment, respectively. For the AFM1 configuration, two
primitive cells consist a magnetic unit cell. In all figures, we
show only the Fe atoms to enhance the
visibility.\label{fig_structure}}
\end{figure*}

\begin{figure*}[htp]
 \centering
 \subfigure[]{
   \rotatebox{270}{\scalebox{0.6}{\includegraphics{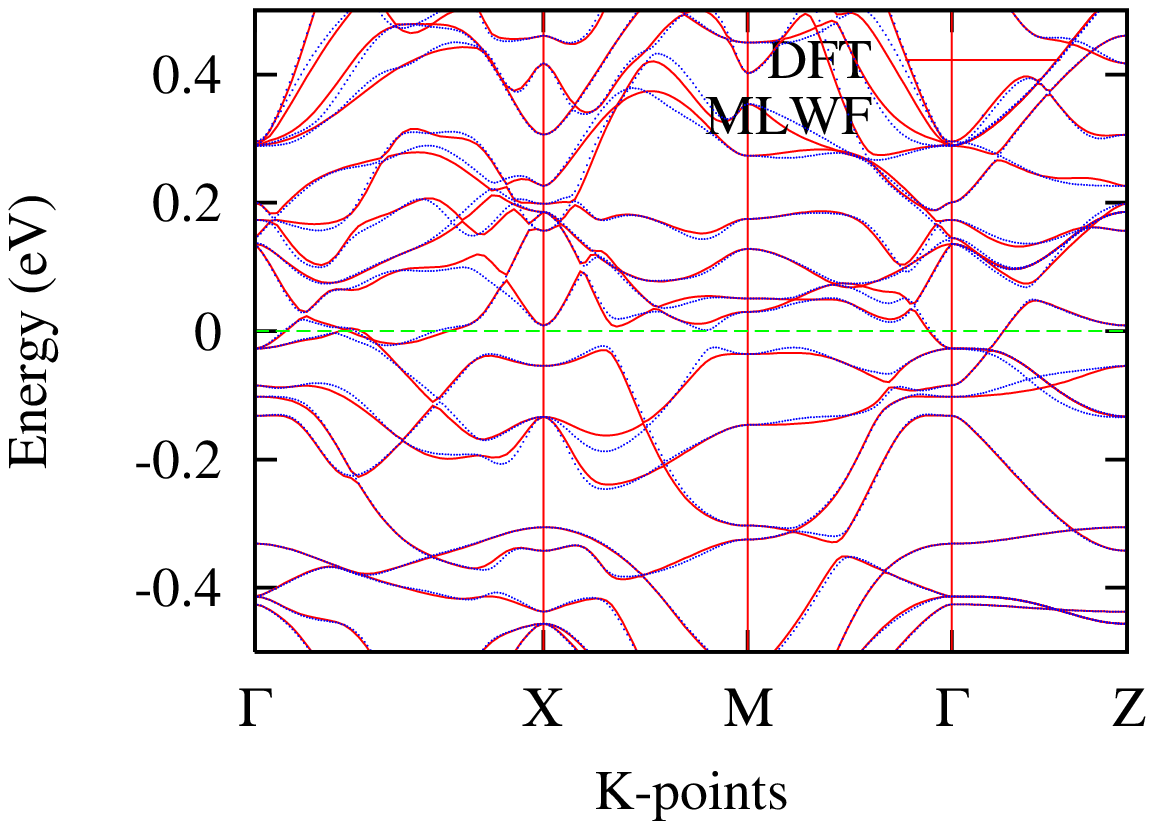}}}
   \label{fig_nm_bs}
 }
 \subfigure[]{
   \rotatebox{270}{\scalebox{0.6}{\includegraphics{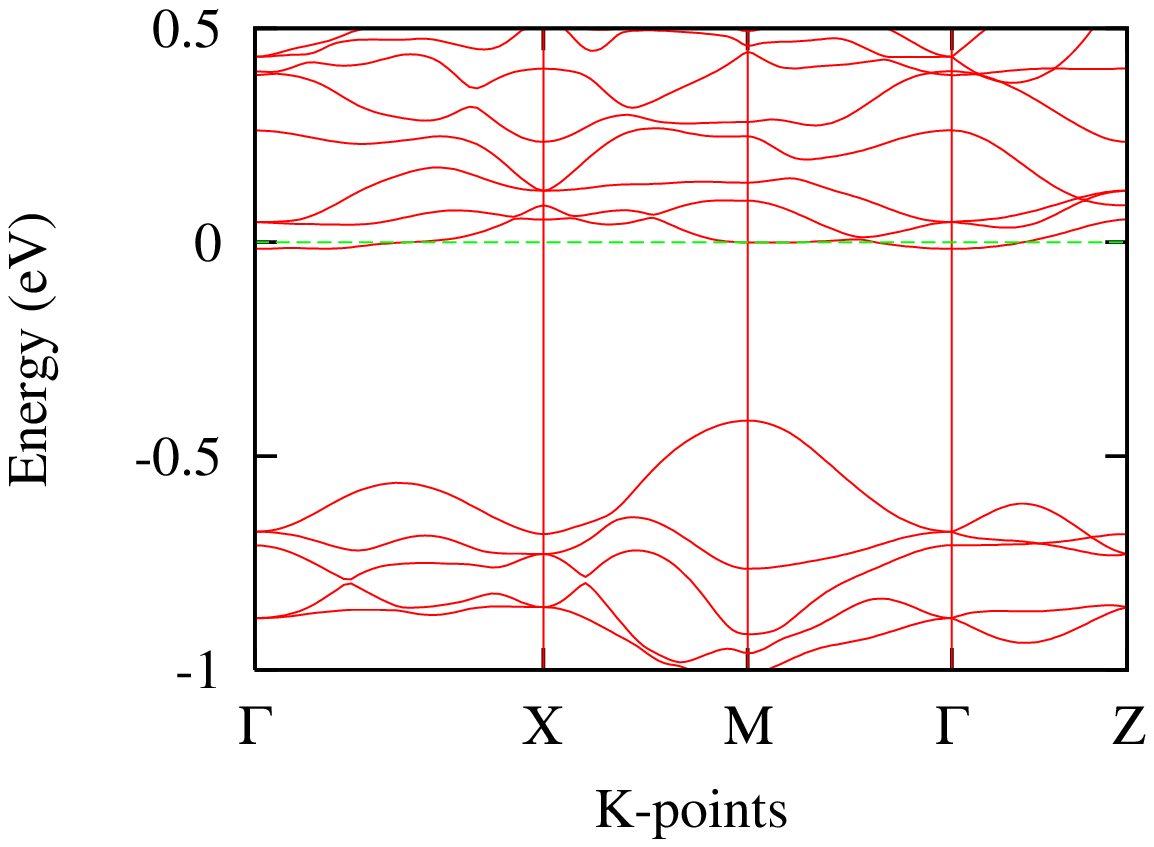}}}
   \label{fig_afm_bs}
 }
 \subfigure[]{
   \rotatebox{270}{\scalebox{0.6}{\includegraphics{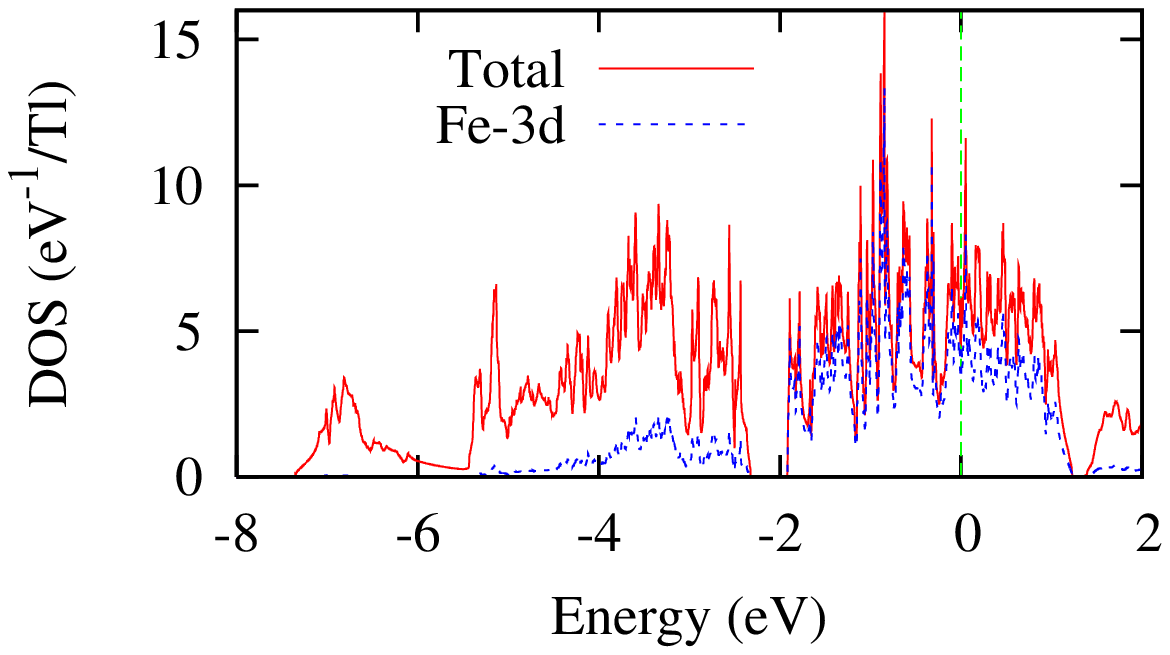}}}
   \label{fig_nm_dos}
 }
 \subfigure[]{
   \rotatebox{270}{\scalebox{0.6}{\includegraphics{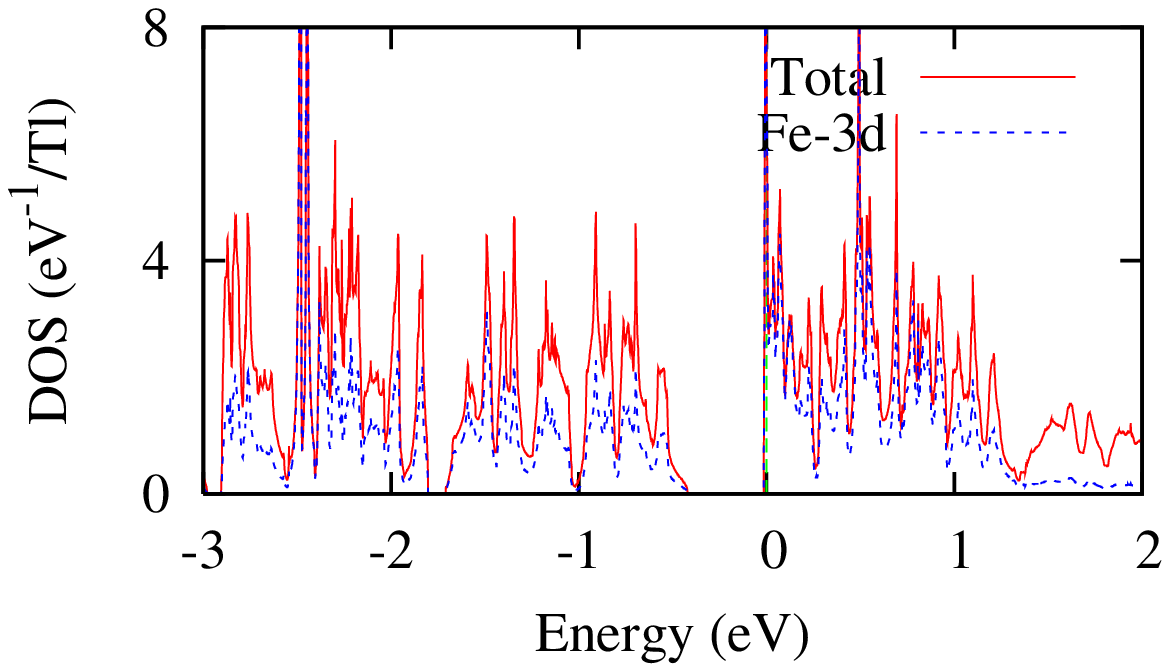}}}
   \label{fig_afm_dos}
 }
 \caption{Band structure and DOS for \tlfsx: The left and right panels are
 for the NM and AFM2 states, respectively. Only up spin is shown in
panels \ref{fig_nm_bs} and \ref{fig_nm_dos} since the up and down
spins are degenerate. \label{fig_bs_dos}}
\end{figure*}

So far it is yet unclear about the precise location of the critical
Fe-deficiency $x_c$ (for $y=1$) where the metal-insulator transition
takes place. One should notice that in experiments the real
Fe-content is sample dependent and may deviate from the nominal
ones, and possibly ( in addition to introduce electrons to the FeSe
layer), the role of Tl atoms is to stablize the Fe-vacancy orderings
while the role of alkaline atoms (K, Rb, Cs) is to achieve higher
Fe-content. Among all the iron deficient compounds, \ktlfs
~($x=0.4$) is of special interest due to its closer proximity to the
transition point $x_c$ and the peculiar ordering pattern of the
Fe-vacancies which can be stablized in tetragonal crystalline
superstructure~\cite{mhfang_1012,arxiv:1101.2059}. This is the
simplest vacancy superstructure with the highest symmetry since all
iron atoms are 3-coordinated equivalently. Thus it is especially
interesting to understand the electronic and magnetic structures of
\ktlfs with the tetragonal Fe-vacancy superstructure.

In this paper, we performed extensive study on \tlfs, using the
first-principles simulations. We have also performed calculations on
KFe$_{1.6}$Se$_2$ for the magnetic structure as well as the density
of states (DOS), which agrees well with \tlfs\ results. We found
that tuning $y$ only leads to the rigid band shift for a specific
Fe-deficiency $x$. Thus our results are valid for the mixture
system \ktlfs. In particular, we used the Vienna Ab-initio
Simulation Package (VASP) \cite{vasp_1,vasp_2}, which employs the
plane-wave basis set and the projected augmented wave (PAW) method
\cite{bloch_paw}. A body-centered orthorhombic primitive cell (FIG.
\ref{fig_structure_geometry}) was used throughout the calculation
unless otherwise specified. A 360 eV energy cut-off and a
$4\times4\times4$ $\Gamma$-centered k-grid were chosen to ensure the
convergence of the total energy to 1 meV/cell. All the geometry were
optimized until the forces on each atom smaller than 0.01 eV/\AA\
and the total pressure smaller than 0.5 kB. For the DOS
calculations, a much finer k-grid of $16\times16\times16$ and the
tetrahedra method were used.

Similar to (K,Tl)Fe$_{1.5}$Se$_2$, \ktlfs~ has three different
stacking patterns. Here we focus on the in-plane magnetic structure
and consider the $AA$-stacking only. As indicated in
(K,Tl)Fe$_{1.5}$Se$_2$, the stacking ordering contributes only a
negligible secondary correction to the total energy unless it
changes the symmetry of the crystal lattice\cite{arxiv:1101.0533}.
We also consider the anti-ferromagnetic (AFM) inter-layer coupling
while its magnitude is negligible owing to the large inter-layer
distances. As a check, we performed test calculations on one of our
spin configurations. It turns out that inter-layer magnetic coupling
contribution is $<$ 1 meV/Fe for \tlfs.

\begin{table}[htp]
 \caption{Lattice constants and magnetic properties of \tlfs. The lattice constants are transformed to represent 122 crystal.
 }
 \label{tab_geo_energies}
 \begin{tabular}{c|c|c|c|c}
     & $a(b)$ (\AA) & $c$ (\AA) & $m_{\mathrm{Fe}}$ ($\mu_B$) & $E_{\Delta}$ (meV/Fe) \\
  \hline\hline
  NM   & 3.8649(3.8649) & 13.1812 & 0 & 0 \\
  FM   & 3.7903(3.7904) & 14.7087 & 2.8 & -62 \\
  AFM0 & 3.9026(3.9026) & 13.6949 & 2.3 & -131 \\
  AFM1 & 3.8494(3.8182) & 14.1779 & 2.7 & -183 \\
  AFM2 & 3.8667(3.8668) & 14.2420 & 2.8 & {\bf -254} \\
  AFM3 & 3.8043(3.8517) & 14.1884 & 2.7 & -199 \\
  AFM4 & 3.8892(3.8892) & 13.9058 & 2.5 & -175 \\
  AFM5 & 3.8882(3.8778) & 13.9053 & 2.6 & -180 \\
  AFM6 & 3.7645(3.8440) & 14.3045 & 2.8 & -200 \\
  AFM7 & 3.8994(3.8385) & 14.0930 & 2.6 & -183 \\
 \end{tabular}
\end{table}

To begin with, we first study the ground state magnetism by
considering 8 possible in-plane AFM configurations [FIG.
\ref{fig_afm0} to \ref{fig_afm7}], as well as the non-magnetic (NM)
and ferro-magnetic (FM) orderings. We list the relaxed geometry
parameters as well as their relative energies in TABLE
\ref{tab_geo_energies}. The AFM0 configuration could be regarded as
the checkerboard AFM; whereas the AFM1 and AFM6 configurations are
the bi-collinear and zig-zag collinear orderings respectively. Our
calculations suggest a ground state of the AFM2 type, whose
configuration energy is 433 meV/cell (or 54 meV/Fe) lower than the
second lowest (AFM6) configuration. Due to the symmetry of
Fe-vacancies, all Fe sites are equivalent in \ktlfs, forming perfect
square Fe blocks (indicated by the blue units in FIG.
\ref{fig_mag_couple}) intercalated by Fe-vacancies. The AFM2
configuration can thus be regarded as checkerboard
antiferromagnetically coupled blocks of parallel aligned spins.
Furthermore, due to the structure distortion induced by the
Fe-vacancies, the magnetic couplings could be classified into two
groups: the ones within a square block (intra-block) and the ones
across two near-by square blocks (inter-block). To model our
energetic results, we incorporated a spin model involving both the
nearest neighbour (n.n.) and the next nearest neighbour (n.n.n.)
couplings:
\begin{equation}
 \begin{split}
H=&\sum_{\substack{n,\alpha}}(J_1 S_{n,\alpha} S_{n,\alpha+1}+J'_1 S_{n,\alpha_{\delta}} S_{n+\delta,\alpha_{\delta}})+\\
 & \sum_{\substack{n,\alpha}}J_2 S_{n,\alpha} S_{n,\alpha+2}+ \\
 & \sum_{\substack{n,\alpha}}J'_2 (S_{n,\alpha_{\delta}} S_{n+\delta,\alpha_{\delta}+1}+S_{n,\alpha_{\delta}-1} S_{n+\delta,\alpha_{\delta}}).\end{split}
 \label{eq_jj_model}
\end{equation}
Where, $n$ denotes the block index, $\delta$ is short for the
nearest neighbouring block to block $n$, $\alpha$ is the site-index
which goes from 1 to 4, $\alpha_{\delta}$ selects the site
connecting to the nearest neighbouring block $\delta$; $J_1$ and
$J'_1$ ($J_2$ and $J'_2$) indicate the n.n. (the n.n.n.)
couplings of intra- and inter-block, respectively. If we further use
the approximation that only the $S^z$ component is involved (Ising
model with $S^z$ being the same in all configurations at each Fe
sites), we could fit the energetics of the 8 AFM configurations
using the least squares method to obtain $\tilde{J}=2J S^2$. The
resulting intra-block couplings $\tilde{J_1}$ and $\tilde{J_2}$ are
-86 meV and -9 meV, respectively; while the inter-block couplings
$\tilde{J'_1}$ and $\tilde{J'_2}$ are -29 meV and 38 meV,
respectively, with fitting correlation $\sigma=97.04\%$. The AFM
inter-block n.n.n. interaction $\tilde{J'_2}$ dominates the inter-block
interactions and both intra-block interactions are FM, thus a block-type checkerboard AFM configuration is
favored.


We then examine the electronic structure of \ktlfs. We show the band
structures of the AFM2 and NM states for $y$=1.0 in FIG. \ref{fig_afm_bs} and
\ref{fig_nm_bs}, respectively.  The band structure of the AFM2 state
suggests a metallic nature, with a band gap $\sim$ 400 meV  for
TlFe$_{1.6}$Se$_2$ (or 550 meV for KFe$_{1.6}$Se$_2$) for the states
16 meV (or 39 meV ) below $E_F$.
Interestingly, the DOS from the top of the band gap to $E_F$ integrates
to exactly 0.2 electron per \ktlfs\ ($y$=1.0) formula, suggesting that the
material would become an insulator if (K,Tl) content is decreased by
20\%. As a check, we have further performed calculations for
K$_{0.8}$Fe$_{1.6}$Se$_2$, assuming two types of K-vacancy
orderings\footnote{There are two inequivalent K sites, those directly above 
a Fe$_4$ block and others. The two K-vacancy orderings we assumed refer to 
the vacancies occupy either of the two sites.}. Then a band gap $\sim 600$ meV shows up at $E_F$ for
either cases. The result implies that the K-vacancies shift the
chemical potential but their orderings do not change the band
structures.

Another interesting feature of the \ktlfs ~electronic structure is
that in the NM state the band energies disperse significantly along
$k_z$-axis, manifesting its highly 3-dimensional characteristics.
This feature is fundamentally different from the KFe$_2$Se$_2$
compound\cite{cao_dai} and all other iron pnictides. Of course, both
the NM and AFM DOS indicate that the Fe-3d orbitals dominate the
states near $E_F$, similar to all iron-based superconductors.

\begin{figure}[htp]
 \centering
 \subfigure[]{
  \scalebox{0.21}{\includegraphics{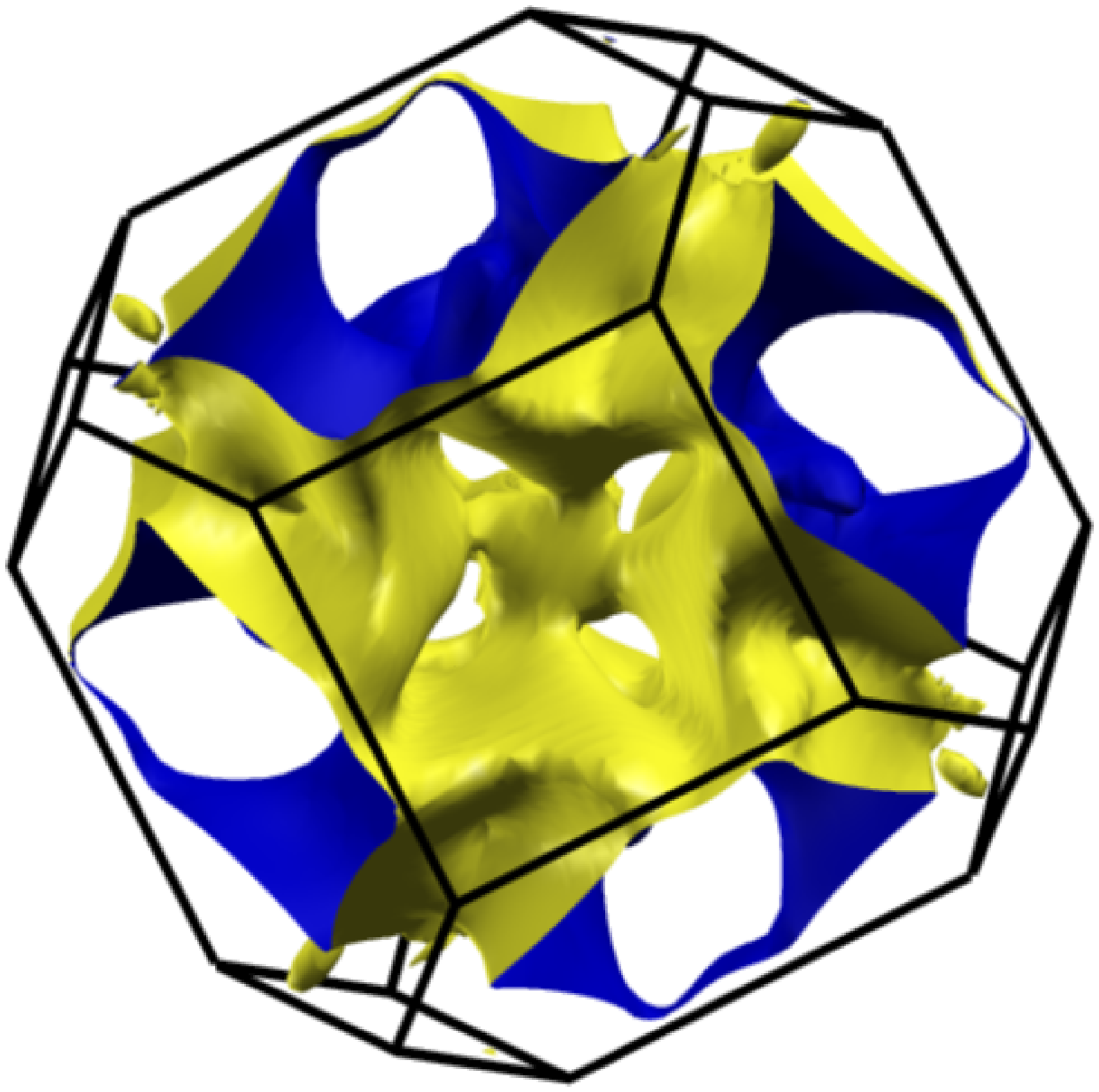}}
  \label{fig_fs_1}
 }
 \subfigure[]{
  \scalebox{0.2}{\includegraphics{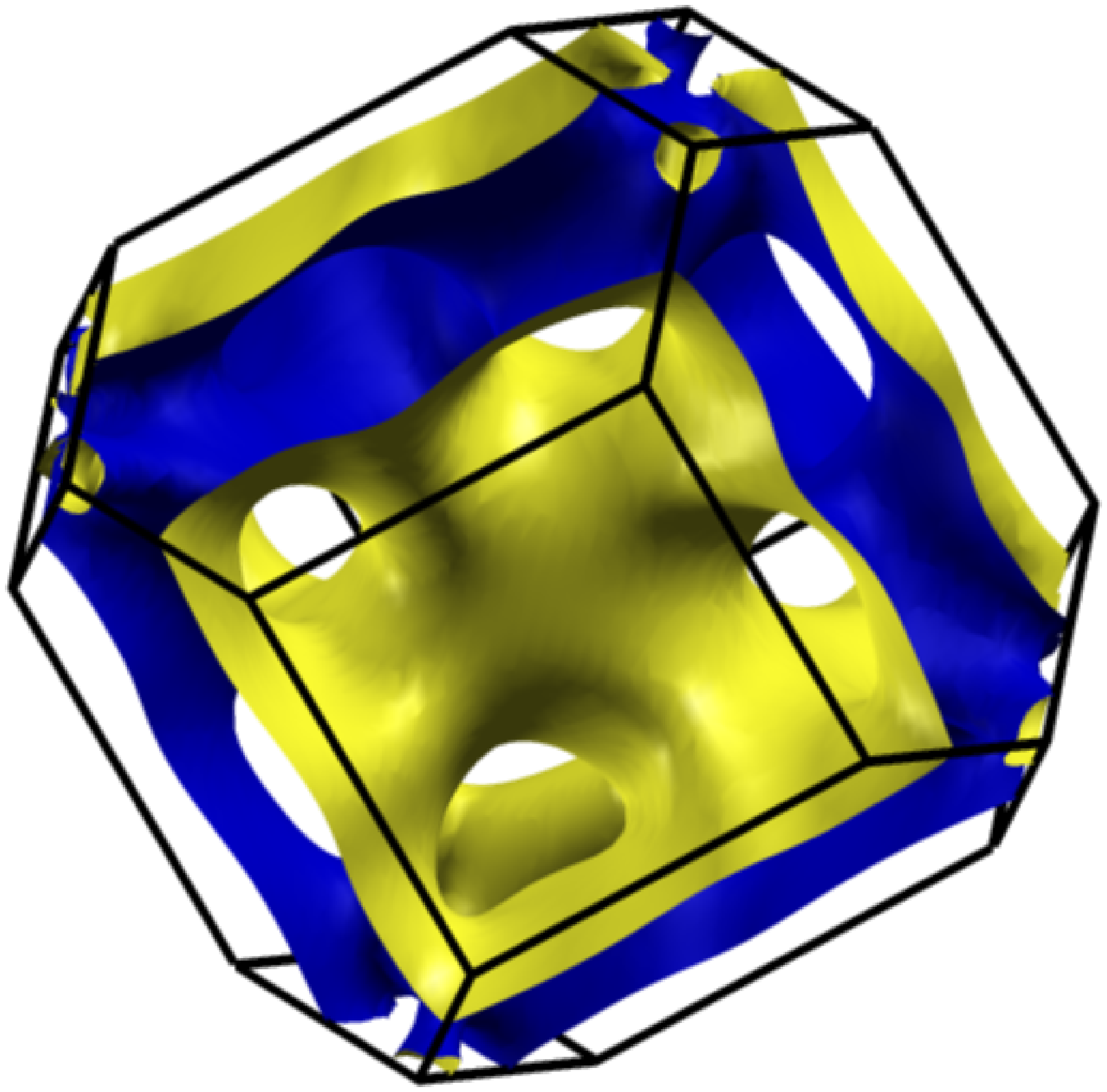}}
  \label{fig_fs_2}
 }
 \subfigure[]{
  \rotatebox{270}{\scalebox{0.5}{\includegraphics{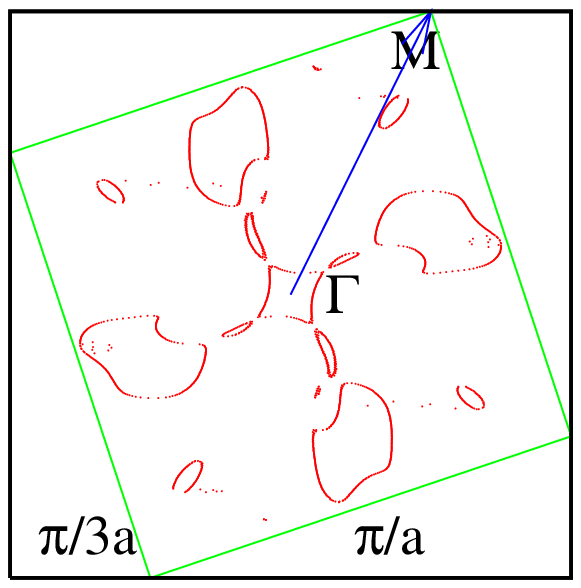}}}
  \label{fig_fs_kz_0_nm}
 }
 \subfigure[]{
  \rotatebox{270}{\scalebox{0.5}{\includegraphics{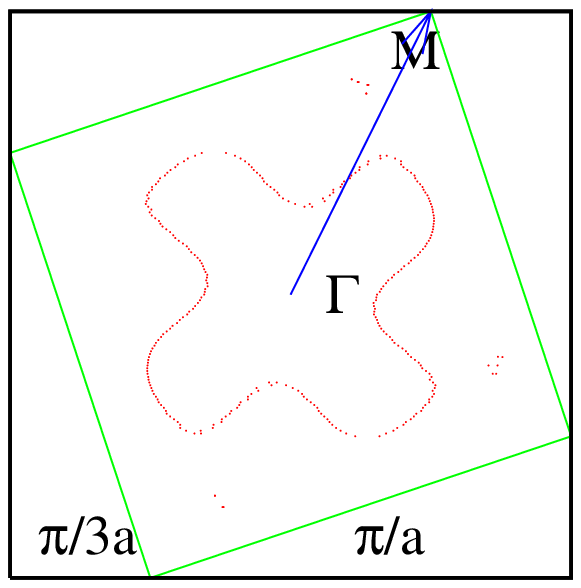}}}
  \label{fig_fs_kz_0_afm}
 }
 \caption{Fermi surfaces of \tlfs reconstructed using MLWFs at
(a,c): non-magnetic (NM) state and (b,d): block-spin anti-ferromagnetic
state; (c,d): Cross section of the Fermi surface of \tlfsx at $k_z=0.0$
plane with $x=0.4$. Both NM and block spin AFM Fermi surfaces are 3D-like.\label{fig_fs}}
\end{figure}

Finally, we reconstruct the Fermi surfaces (FIG. \ref{fig_fs}) of
the metallic \ktlfs\ ($y$=1.0) by fitting the band structure using the
maximally localized wannier functions (MLWFs)\cite{mlwf_1,mlwf_2}.
The Fermi surface of both NM and block-spin AFM \ktlfs\ are highly
3- dimensional, although their specific shape is quite different
from each other. It is worthwhile to notice that due to the
Fe-vacancy superstructure the first Brillouin zone (BZ) is not the
same as the one in (K,Tl)Fe$_2$Se$_2$, and thus the M point in FIG.
\ref{fig_fs} is not ($\pi/a$,$\pi/a$). Nevertheless, the Fermi
surface topology of the present \ktlfs~ compound is quite unique
compared to either KFe$_2$Se$_2$\cite{cao_dai} or all other iron
pnictides. It strongly indicates that the formation of Fe-vacancy
superstructures is crucial to the electronic structures of the
\ktlfsx~ compounds. Unlike the change in $y$, the topological change
of the Fermi surface across $x\sim 0.4$ indicates that the
electronic and magnetic structures for different Fe-vacancy ordered
materials can not be approached by a rigid shift of the chemical
potential.

It is worth noting that while the Fe-vacancy ordering is crucial to
the block spin magnetic pattern and the finite gap for
(K,Tl)$_{0.8}$Fe$_{1.6}$Se$_2$, the strong disorder of the
Fe-vacancies may destroy the magnetic ordering leading to the
metallic ground state. In the vacancy disordered state, the first
Brillouin Zone would remain the same as (K,Tl)Fe$_2$Se$_2$ and thus
the rigid-band model would remain effective. The whole electronic
structure in that case could then be approximated with a hole-doped
(K,Tl)$_y$Fe$_2$Se$_2$\cite{cao_dai}. Actually, it has been
suggested that randomly distributed Fe-vacancies in the doped
antiferromagnetic (or Mott) insulator may lead to a spin-singlet
s-wave superconductor \cite{ZhouZhang}.

In conclusion, we have performed first-principles calculations on
\ktlfs. A block-type checkerboard antiferromagnetic ground state was
identified and the AFM inter-block n.n.n. coupling interaction
dominates. Our calculations suggest a metallic ground state for
$y=1$ with a 400-550 meV band gap which appears slightly below the
Fermi level and an insulating ground state for
(K,Tl)$_{0.8}$Fe$_{1.6}$Se$_2$. The experimentally observed
insulating behavior may be due to both the 20\% (K,Tl) deficiency
and the iron vacancy superstructure. Furthermore, the electronic
structures of the metallic states show a significant 3-dimensional
feature with a unique Fermi surface topology, indicating that the
formation of Fe-vacancy superstructure is crucial to the physical
properties of \ktlfsx~.

The authors would like to thank Q. Si, M.H. Fang and H.D. Wang for helpful
discussions. All calculations were performed at the High Performance
Computing Center of Hangzhou Normal University College of Science.
This work was supported by the NSFC, the NSF of Zhejiang Province,
the 973 Project of the MOST and the Fundamental Research Funds for
the Central Universities of China (No. 2010QNA3026).

{\it Note added}: Recently, we become aware of a paper by W. Bao {\it et al.}
\cite{arXiv:1102.0830} on neutron diffraction
experiment for K$_{0.8}$Fe$_{1.6}$Se$_2$. Our result is in agreement
with the reported magnetic ordering pattern.

\bibliographystyle{apsrev}
\bibliography{tl122}
\end{document}